\documentclass[aps,preprint]{revtex4}%
\usepackage{amsfonts}
\usepackage{amsmath}
\usepackage{amssymb}
\usepackage{graphicx}%
\begin{document}
\title{Cold dilute neutron matter on the lattice I: \ Lattice virial coefficients and
large scattering lengths}
\author{Dean Lee and Thomas Sch{\"a}fer}
\affiliation{Department of Physics, North Carolina State University, Raleigh, NC 27695}
\keywords{nuclear lattice simulation non-perturbative chiral effective field theory}
\pacs{21.30-x,21.65+f,13.75.Cs}

\begin{abstract}
We study cold dilute neutron matter on the lattice using an effective field
theory. \ We work in the unitary limit in which the scattering length is much
larger than the interparticle spacing. \ In this paper we focus on the
equation of state at temperatures above the Fermi temperature and compare
lattice simulations to the virial expansion on the lattice and in the
continuum. \ We find that in the unitary limit lattice discretization errors
in the second virial coefficient are significantly enhanced. \ As a
consequence the equation of state does not show the universal scaling behavior
expected in the unitary limit. \ We suggest that scaling can be improved by
tuning the second virial coefficient rather than the scattering length.

\end{abstract}
\maketitle


\section{Introduction}

\label{sec_int}

Cold dilute neutron matter is an intriguing physical system. \ It is relevant
to the physics of the inner crust of neutron stars \cite{Pethick:1995di}. \ It
is also close to an interesting universal limit which is the result of a large
separation of scales. \ The neutron scattering length is $a_{nn}\simeq-18$ fm,
which implies that the dimensionless parameter $k_{F}|a_{nn}|\gg1$ for
densities $\rho>10^{-4}\rho_{N}$. \ Here, $k_{F}=(3\pi^{2}\rho)^{1/3}$ is the
Fermi momentum and $\rho_{N}\simeq0.16$ fm$^{-3}$ is the saturation density of
nuclear matter. \ The effective range, on the other hand, is $r_{nn}\simeq2.8$
fm. \ So if the density is very small, $\rho<0.1\rho_{N}$, then $k_{F}%
|r_{nn}|$ is a small parameter and neutron matter is close to the limit in
which $k_{F}|a_{nn}|\rightarrow\infty$ and $k_{F}|r_{nn}|\rightarrow0$. \ This
is known as the unitary limit, where the vacuum scattering amplitude in the
s-wave channel has a zero-energy resonance, and the cross-section saturates
the unitarity bound. \ In this limit there is no expansion parameter and the
calculation of the equation of state and of transport properties is a
difficult non-perturbative problem. \ It is now possible to create systems of
fermions in the unitary limit in the laboratory by trapping fermionic atoms
and tuning the scattering length using a Feshbach resonance
\cite{O'Hara:2002,Gupta:2002,Regal:2003,Bourdel:2003,Gehm:2003}.  This
technique will provide experimental measurements of universal parameters, but
it is clearly desirable to also develop a computational approach. \ There are
several computational studies of neutron matter at zero temperature using
potential models and Green's function Monte Carlo
\cite{Carlson:2003wm,Carlson:2003z,Chang:2004sj,Pederiva:2004iz}. \ Recently,
there have also been simulations on the lattice using effective field theory
\cite{Muller:1999cp,Lee:2004si,Lee:2004qd,Wingate:2005xy,Bulgac:2005a}. \ The
main advantage of the effective field theory/lattice approach for the dilute
neutron matter problem is that it is not restricted to zero temperature.

We have described our method and some initial results in \cite{Lee:2004qd}.
\ This is the first in a sequence of two papers in which we perform a careful
study of the parameter dependence and scaling behavior of the results. \ We
are particularly interested in verifying that the lattice results satisfy the
scaling relations that are expected to hold in the unitary limit. \ The main
focus of this first paper is to understand the equation of state at low
density and high temperature and compare the results to the virial expansion.
\ We find that if the scattering length is large lattice discretization errors
can be as large as $100\%$ or more. \ At fixed lattice spacing when the
temperature decreases we find that the error first increases before eventually
decreasing. \ In order to counter this effect, we propose that the interaction
coefficient be tuned to give the correct second order virial coefficient,
$b_{2}(T)$, at the chosen simulation temperature. \ We describe in detail how
this is done. \ In the sequel to this paper we apply this technique to
simulations of cold dilute neutron matter in the unitary limit and study the
scaling behavior of the results.


\section{Lattice Action}

\label{sec_act}

We consider the theory defined by the partition function
\begin{equation}
Z_{G}=Tr\exp\left[  -\beta(H-\mu N)\right]  \simeq z_{0}e^{2\mu\beta L^{3}%
}\int DsDc^{\prime}Dc^{\ast}\exp\left[  -S\right]  ,
\end{equation}
where the lattice action is given by
\begin{align}
S  &  =\sum_{\vec{n},i}\left[  e^{-\mu\alpha_{t}}c_{i}^{\ast}(\vec{n}%
)c_{i}^{\prime}(\vec{n}+\hat{0})-e^{\sqrt{-C\alpha_{t}}s(\vec{n}%
)+\frac{C\alpha_{t}}{2}}(1-6h)c_{i}^{\ast}(\vec{n})c_{i}^{\prime}(\vec
{n})\right] \nonumber\\
&  -h\sum_{\vec{n},l_{s},i}\left[  c_{i}^{\ast}(\vec{n})c_{i}^{\prime}(\vec
{n}+\hat{l}_{s})+c_{i}^{\ast}(\vec{n})c_{i}^{\prime}(\vec{n}-\hat{l}%
_{s})\right]  +\frac{1}{2}\sum_{\vec{n}}s^{2}(\vec{n}).
\end{align}
Here, $\vec{n}$ labels the sites of a $3+1$ dimensional lattice, $\hat{l}%
_{s}\ (s=1,2,3)$ is a spatial unit vector, and $i$ labels the two spin
components of the neutron, $\uparrow$ and $\downarrow$. \ The spatial lattice
spacing is $a$ and $\alpha_{t}=a_{t}/a$ is the ratio of the temporal to the
spatial lattice spacing. \ The spatial volume of the lattice is $L^{3}$ and
the temporal length is $\beta=1/T$. \ Dimensional quantities like the chemical
potential $\mu$ and the nucleon mass $m_{N}$ are given in units of the lattice
spacing $a$. \ We have also defined $h=\alpha_{t}/(2m_{N})$. The Grassmann
fields are denoted by $c_{i}(\vec{n})$ and $s(\vec{n})$ is a
Hubbard-Stratonovich field. \ 

\begin{figure}
[ptb]
\begin{center}
\includegraphics[
height=0.8856in,
width=2.284in
]%
{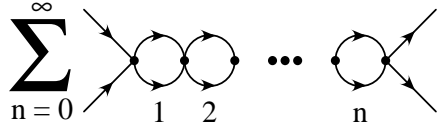}%
\caption{Diagrams contributing to neutron-neutron scattering.}%
\label{scattering}%
\end{center}
\end{figure}

The interaction coefficient $C$ must be determined for given lattice spacings
$a$ and $a_{t}$. \ In \cite{Lee:2004qd} this was done by adjusting $C$ to
reproduce the correct neutron scattering length at zero temperature. \ This
requires summing the two-particle scattering diagrams shown in Fig.
\ref{scattering}.
%
The pole in the scattering amplitude is then compared with L\"{u}scher's
formula for energy levels in a finite periodic box
\cite{Luscher:1986pf,Beane:2003da},%
\begin{equation}
E_{0}=\dfrac{4\pi a_{\text{scatt}}}{m_{N}L^{3}}[1-c_{1}\frac{a_{\text{scatt}}%
}{L}+c_{2}\frac{a_{\text{scatt}}^{2}}{L^{2}}+\cdots], \label{lus}%
\end{equation}
where $c_{1}=-2.837297,$ $c_{2}=6.375183$.


\section{Results for large scattering lengths}

\label{sec_eos}

We present lattice simulation results for the energy per particle as a
function of density for several different scattering lengths. \ We are
interested in cold dilute neutron matter where all length scales are much
larger than the lattice spacing. \ We use a spatial lattice spacing of $a=(50$
MeV$)^{-1}$ and temporal lattice spacing of $a_{t}=(24$ MeV$)^{-1}$. \ We will
consider scattering lengths $|a_{\text{scatt}}|>4.675$ fm. \ The spatial
lattice spacing was chosen so that it is smaller than the smallest scattering
length. \ The temporal lattice spacing was chosen sufficiently small so that
the results are very close to the $a_{t}\rightarrow0$ limit.\ We have computed
the energy per neutron for five different scattering lengths. \ The
corresponding operator coefficients are shown in Table 1.
\[%
\genfrac{}{}{0pt}{0}{\text{Table 1: \ Operator coefficients and scattering
lengths}}{%
\begin{tabular}
[c]{|l|l|l|l|l|l|}\hline
$C$ ($10^{-4}$ MeV$^{-2}$) & $-1.028$ & $-1.108$ & $-1.153$ & $-1.257$ &
$-1.318$\\\hline
$\frac{dC}{d\alpha_{t}}$ ($10^{-5}$ MeV$^{-2}$) & $-2.259$ & $-2.239$ &
$-2.218$ & $-2.141$ & $-2.078$\\\hline
$a_{\text{scatt}}$ (fm) & $-4.675$ & $-9.35$ & $-18.70$ & $+18.70$ &
$+9.35$\\\hline
\end{tabular}
}%
\]
We have also shown the derivative $\frac{dC}{d\alpha_{t}}$ which is needed to
compute derivatives with respect to $\beta$.

We use the hybrid Monte Carlo algorithm \cite{Duane:1987de} to generate
Hubbard-Stratonovich field configurations as described in \cite{Lee:2004qd}.
\ We use diagonal preconditioning before each conjugate gradient solve. \ If
$K$ is the single-spin neutron matrix for a given Hubbard-Stratonovich field
configuration, then we must solve the linear equation%
\begin{equation}
K^{\dagger}Kv=b\text{.}%
\end{equation}
Rather than solving this directly, we make use of the diagonal matrix%
\begin{equation}
D=\text{diag}\left[  K^{\dagger}K\right]
\end{equation}
and solve instead
\begin{equation}
\left[  D^{-1}K^{\dagger}KD^{-1}\right]  Dv=D^{-1}b.
\end{equation}
Fluctuations associated with the Hubbard-Stratonovich field occur on the
diagonal of $K$, and therefore the matrix $D^{-1}K^{\dagger}KD^{-1}$ typically
has a smaller condition number than $K^{\dagger}K$. \ We further improve the
performance by adding a small positive constant $\epsilon$ to produce the
modified equation%
\begin{equation}
\left[  D^{-1}K^{\dagger}KD^{-1}+\epsilon\right]  Dv=D^{-1}b.
\end{equation}
We tune $\epsilon$ so that the equilibration time for the hybrid Monte Carlo
algorithm is minimized while keeping the induced systematic error smaller than
the stochastic error. \ In practice we take $\epsilon$ to be $10^{-4}$ or smaller.

Roughly $10^{4}$ HMC trajectories were run, split across 4 processors running
completely independent trajectories. \ Averages and errors were computed by
comparing the results of each processor. \ The finite volume error was tested
by going to larger volumes. \ The final lattice sizes were chosen so that the
finite volume error was less than one percent. \ For the data at $T=4$ MeV we
used a lattice of size $4^{3}\times6$.\ For $T=3$ MeV we used $5^{3}\times8$,
and for $T=2$ MeV we used $5^{3}\times12.$ \ As an example of the finite
volume dependence, we show in Table 2 the density and energy per particle from
simulations with lattice volumes $4^{3},5^{3},6^{3}$ at $a_{\text{scatt}%
}=-18.70$ fm, $T=4$ MeV, and $\mu=0$. For comparison we also show the volume
dependence of the bubble chain calculations described in \cite{Lee:2004qd} and
Sect.~\ref{sec_bub}.
\[%
\genfrac{}{}{0pt}{0}{\text{Table 2: }L\text{ dependence for }a_{\text{scatt}%
}=-18.70\text{ fm, }T=4\text{ MeV, }\mu=0}{%
\begin{tabular}
[c]{|l|l|l|l|l|}\hline
$L$ & $\rho^{\text{bubble}}$ ($10^{-3}$ fm$^{-3}$) & $\frac{E^{\text{bubble}}%
}{A^{\text{bubble}}}$(MeV) & $\rho^{\text{sim}}$ ($10^{-3}$ fm$^{-3}$) &
$\frac{E^{\text{sim}}}{A^{\text{sim}}}$(MeV)\\\hline
$4$ & $7.688$ & $3.527$ & $8.70(3)$ & $3.41(3)$\\\hline
$5$ & $7.695$ & $3.521$ & $8.78(10)$ & $3.41(3)$\\\hline
$6$ & $7.695$ & $3.521$ & $8.81(10)$ & $3.39(3)$\\\hline
\end{tabular}
}%
\]
These results suggest that the finite volume effects are smaller than the
statistical errors. \ If we take the volume dependence from the bubble chain
calculations as a guide, then the finite volume errors are well below one percent.

\begin{figure}
[ptb]
\begin{center}
\includegraphics[
height=4.2436in,
width=2.975in,
angle=-90
]%
{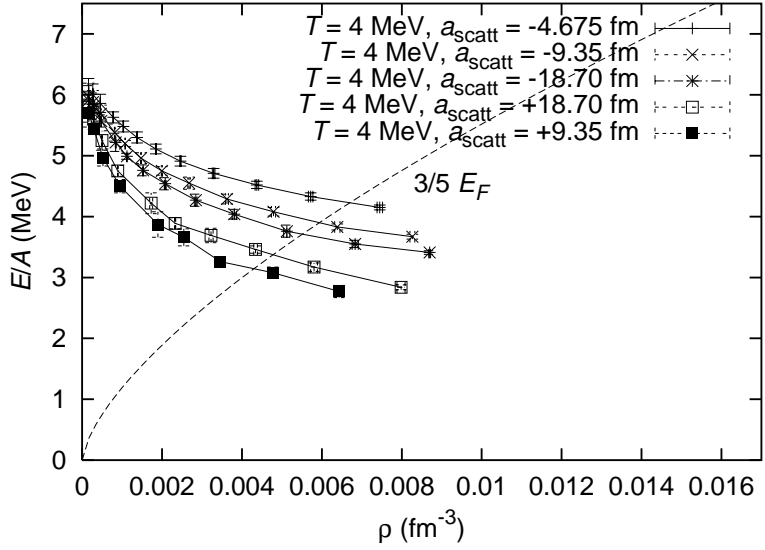}%
\caption{Energy per neutron versus density at $T=4$ MeV for various scattering
lengths.}%
\label{E_A_4}%
\end{center}
\end{figure}

\begin{figure}
[ptbptb]
\begin{center}
\includegraphics[
height=4.2436in,
width=2.975in,
angle=-90
]%
{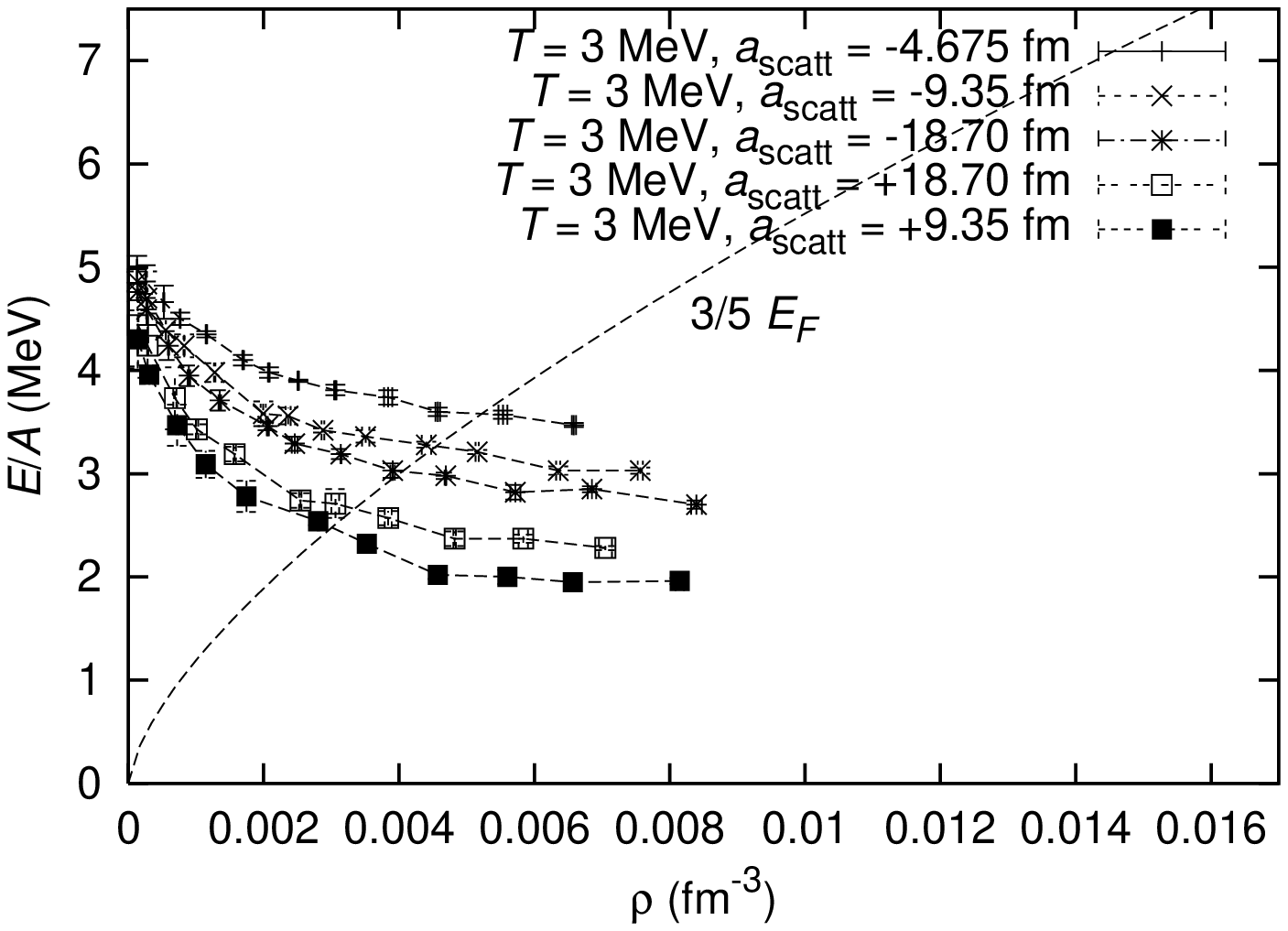}%
\caption{Energy per neutron versus density at $T=3$ MeV for various scattering
lengths.}%
\label{E_A_3}%
\end{center}
\end{figure}

\begin{figure}
[ptbptbptb]
\begin{center}
\includegraphics[
height=4.2436in,
width=2.975in,
angle=-90
]%
{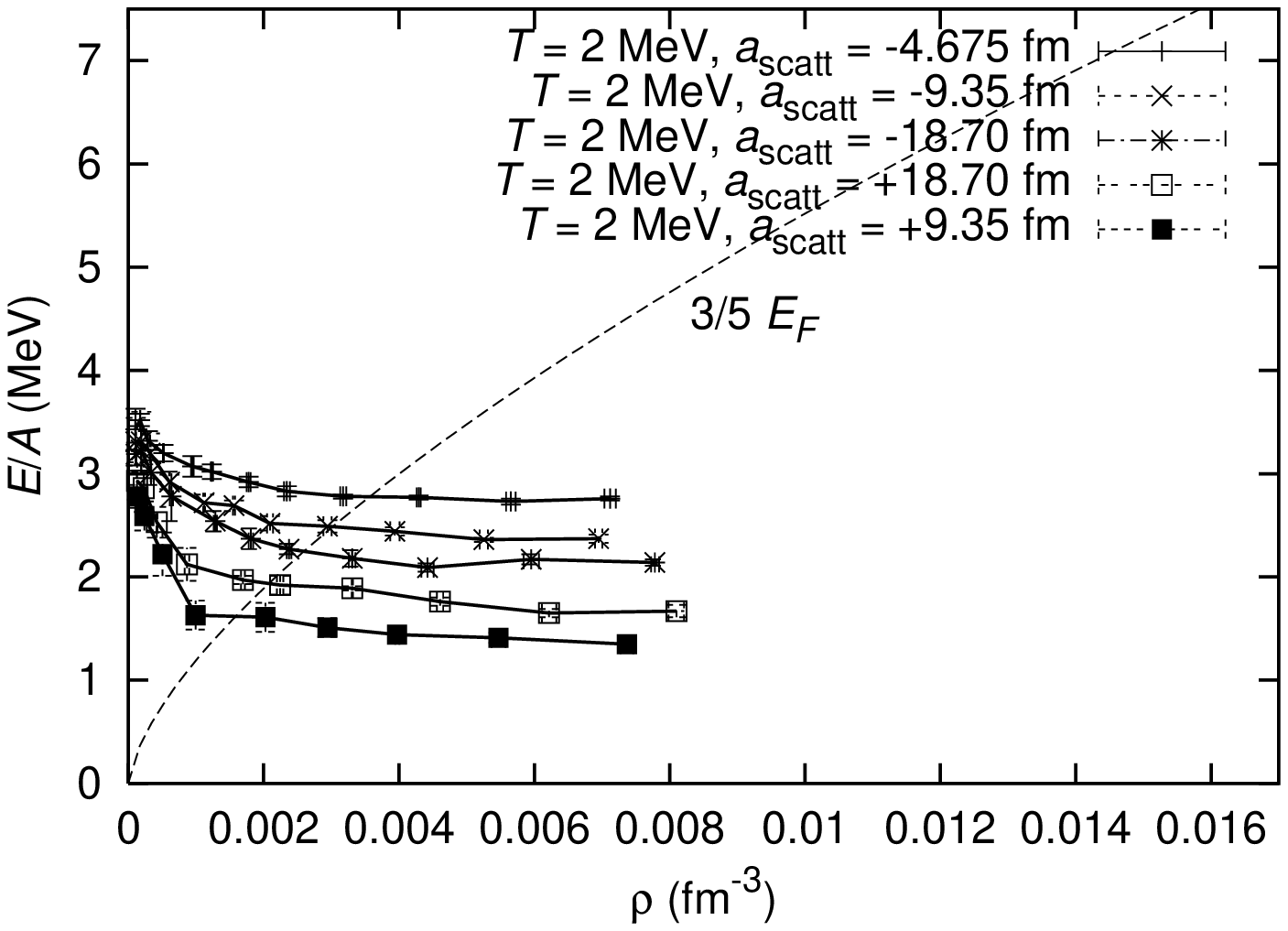}%
\caption{Energy per neutron versus density at $T=2$ MeV for various scattering
lengths.}%
\label{E_A_2}%
\end{center}
\end{figure}

Fig. \ref{E_A_4} shows the lattice simulation results for the energy per
neutron at $T=4$\ MeV. \ Fig. \ref{E_A_3} shows the energy per neutron at
$T=3$\ MeV, and Fig. \ref{E_A_2} shows the energy per neutron at $T=2$\ MeV.
%
%
In each of the plots we have taken a range of densities from zero to a
quarter-filled lattice. \ With a spatial lattice spacing of $(50$ MeV$)^{-1}$,
the quarter-filled lattice corresponds with a density of $0.0081$ fm$^{-3}$.
\ Beyond this one might find significant lattice artifacts. \ We observe that
the energy per particle depends quite strongly on the scattering length, both
at large and at small density. \ While the dependence of $E/A$ on the
scattering length for a degenerate Fermi gas is a complicated,
non-perturbative problem it is possible to compare our results to theoretical
predictions in the opposite limit of low density and high temperature.


\section{Virial expansion}

\label{sec_vir}

The virial expansion arranges multi-particle interactions as a power series in
fugacity,%
\begin{equation}
z=e^{\beta\mu}\text{.}%
\end{equation}
For example the logarithm of the partition function per unit volume can be
written as%
\begin{equation}
\frac{1}{V}\ln Z_{G}=\beta P=\frac{2}{\lambda_{T}^{3}}\left[  z+b_{2}%
(T)z^{2}+b_{3}(T)z^{3}\cdots\right]  .
\end{equation}
The second order virial coefficient $b_{2}(T)$ is determined entirely by
two-particle interactions, while the third virial coefficient depends on
three-body interactions, and so on. \ We can use the virial expansion to
compute thermodynamic observables in the high-temperature/low-density limit.
\ The virial expansion is reliable if the thermal wavelength $\lambda
_{T}=\sqrt{(2\pi)/(m_{N}T)}$ is smaller than the interparticle spacing,
$\lambda_{T}<\rho^{-1/3}$. \ The neutron density can be computed in terms of
the derivative of $\ln Z_{G}$ with respect to the chemical potential,
\begin{equation}
\rho=\frac{A}{V}=\frac{1}{\beta V}\frac{\partial}{\partial\mu}\ln
Z_{G}\text{.}%
\end{equation}
To second order in the virial expansion we find%
\begin{equation}
\rho=\frac{2}{\lambda_{T}^{3}}\left[  z+2b_{2}(T)z^{2}+\cdots\right]  .
\label{virialrho}%
\end{equation}
As a result of discretization error we find on the lattice a more general
power series in fugacity,%
\begin{equation}
\rho=\frac{2}{\lambda_{T}^{3}}b_{1}(T)\left[  z+2b_{2}(T)z^{2}+\cdots\right]
. \label{lattice virial}%
\end{equation}
While $b_{1}(T)$ is no longer guaranteed to equal $1$, we should find that
$b_{1}(T)\rightarrow1$ in the continuum limit. \ We will use
(\ref{lattice virial}) as the definition for the virial coefficients on the lattice.

In the unitary limit, where the effective range is zero and scattering length
is infinite, one finds \cite{Ho:2004a}%
\begin{equation}
b_{2}(T)=3\cdot2^{-\frac{5}{2}}\approx0.530.
\end{equation}
We can compare this with the result for a free Fermi gas, where%
\begin{equation}
b_{2}(T)=-2^{-\frac{5}{2}}\approx-0.177.
\end{equation}
For zero range but finite scattering length the second virial coefficient
becomes%
\begin{equation}
b_{2}(T)=\left\{
\genfrac{}{}{0pt}{}{\frac{e^{x^{2}}}{\sqrt{2}}\left[  1-\operatorname{erf}%
(\left\vert x\right\vert )\right]  -\frac{1}{4\sqrt{2}}\text{ \ for
}x<0,}{\sqrt{2}e^{\frac{\left\vert E_{B}\right\vert }{k_{B}T}}-\frac{e^{x^{2}%
}}{\sqrt{2}}\left[  1-\operatorname{erf}(x)\right]  -\frac{1}{4\sqrt{2}}\text{
\ for }x>0,}%
\right.
\end{equation}
where $\operatorname{erf}$ is the error function, $E_{B}$ is the two-particle
bound state energy for positive scattering length, and%
\begin{equation}
x=\frac{\lambda_{T}}{\sqrt{2\pi}a_{\text{scatt}}}.
\end{equation}
As the effective range goes to zero we have the relation
\begin{equation}
\left\vert E_{B}\right\vert =\frac{1}{ma_{\text{scatt}}^{2}},
\end{equation}
and therefore we can write%
\begin{equation}
b_{2}(T)=\left\{
\genfrac{}{}{0pt}{}{\frac{e^{x^{2}}}{\sqrt{2}}\left[  1-\operatorname{erf}%
(\left\vert x\right\vert )\right]  -\frac{1}{4\sqrt{2}}\text{ \ for
}x<0,}{\sqrt{2}e^{x^{2}}-\frac{e^{x^{2}}}{\sqrt{2}}\left[
1-\operatorname{erf}(x)\right]  -\frac{1}{4\sqrt{2}}\text{ \ for }x>0.}%
\right.  \label{b2}%
\end{equation}
See \cite{Horowitz:2005zv} for a calculation of $b_{2}(T)$ with realistic
interactions, taking into account effective range corrections as well as
higher partial waves.


\section{Bubble chain diagrams and the virial expansion on the lattice}

\label{sec_bub}

\begin{figure}
[ptb]
\begin{center}
\includegraphics[
height=1.1744in,
width=2.284in
]%
{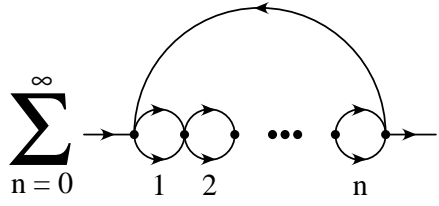}%
\caption{Bubble chain diagrams contributing to the neutron self-energy.}%
\label{self}%
\end{center}
\end{figure}

\begin{figure}
[ptbptb]
\begin{center}
\includegraphics[
height=1.4252in,
width=2.0435in
]%
{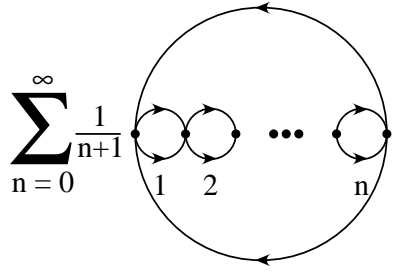}%
\caption{Bubble chain diagrams contributing to the logarithm of the partition
function.}%
\label{bubble}%
\end{center}
\end{figure}

The second virial coefficient is determined by two body interactions. \ This
means that we do not have to rely on simulations in order to determine
$b_{2}(T)$ on the lattice, but we can extract the second virial coefficient
from the lattice regularized bubble chain diagrams. \ For the neutron
propagator we have the diagrams shown in Fig. \ref{self}.
%
The bubble diagrams give a contribution to neutron self-energy of the form
\cite{Lee:2004qd}%
\begin{equation}
\Sigma(\vec{q})=-(1-6h)^{2}\left(  e^{-C\alpha_{t}}-1\right)  \frac{1}%
{L^{3}L_{t}}\sum_{\vec{p}}\frac{D^{\text{free}}(\vec{p}-\vec{q})}%
{1-(1-6h)^{2}\left(  e^{-C\alpha_{t}}-1\right)  B(\vec{p},\mu)}%
\end{equation}
where%
\begin{equation}
B(\vec{p},\mu)=\frac{1}{L^{3}L_{t}}\sum_{\vec{k}}\frac{1}{e^{-\mu\alpha_{t}%
}e^{-ip_{\ast0}/2}e^{-ik_{\ast0}}-1+\omega_{p/2+k}}\frac{1}{e^{-\mu\alpha_{t}%
}e^{-ip_{\ast0}/2}e^{ik_{\ast0}}-1+\omega_{-p/2+k}}.
\end{equation}
We use this to compute the full neutron propagator%
\begin{equation}
D^{\text{full}}(\vec{q})=\frac{D^{\text{free}}(\vec{q})}{1-\Sigma(\vec
{q})D^{\text{free}}(\vec{q})}.
\end{equation}
The average number of neutrons is then%
\begin{equation}
A=\frac{1}{\beta}\frac{\partial}{\partial\mu}\ln Z_{G}=2L^{3}\left[
1-\frac{e^{(m_{N}-\mu)\alpha_{t}}}{L_{t}L^{3}}\sum_{\vec{k}}D^{\text{full}%
}(\vec{k})e^{-ik_{\ast0}}\right]  .
\end{equation}
For the logarithm of the grand canonical partition function, $\ln Z_{G}$, the
lowest non-trivial order in $\rho\lambda_{T}^{3}$ is given by the bubble chain
diagrams shown in Fig. \ref{bubble}.
%
These give a contribution%
\begin{align}
&  \ln Z_{G}-\ln Z_{G}^{\text{free}}\nonumber\\
&  =\frac{1}{L_{t}L^{3}}\sum_{\vec{p},\vec{q}}\frac{-\ln\left[  1-(1-6h)^{2}%
\left(  e^{-C\alpha_{t}}-1\right)  B(\vec{p}+\vec{q},\mu)\right]
D^{\text{free}}(\vec{p})D^{\text{free}}(\vec{q})}{B(\vec{p}+\vec{q},\mu)}.
\end{align}

\begin{figure}
[ptb]
\begin{center}
\includegraphics[
height=3.6288in,
width=3.3935in,
angle=-90
]%
{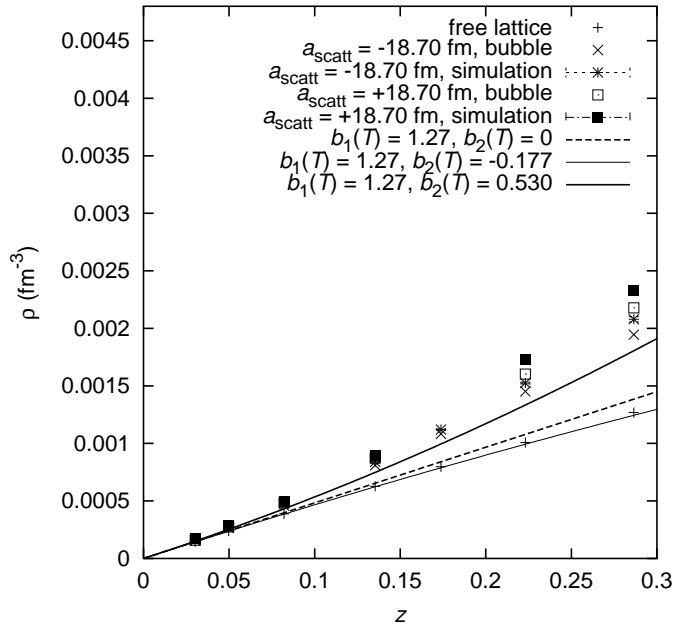}%
\caption{Density versus fugacity at $T=$ $4$\ MeV for scattering lengths
$a_{\text{scatt}}=\pm18.70$ fm. \ We show comparisons with the first and
second order virial expansion.}%
\label{rho_z_4}%
\end{center}
\end{figure}

We can now compare these results to the results from simulations and to the
virial expansion in the continuum. \ In Fig. \ref{rho_z_4} we plot the density
versus fugacity at $T=4$ MeV for scattering lengths $a_{\text{scatt}}=
\pm18.7$ fm$.$ \ We show data for a free Fermi gas on the lattice, bubble
chain calculations, and full simulation results. \ We also plot first and
second order virial results using $b_{1}(T)=1.27$, which we obtained by
fitting the bubble chain data at very small fugacity. \ The results for the
free Fermi gas on the lattice agree with the second order virial results with
$b_{2}(T)=-0.177$. \ The bubble chain and full simulation results for
$a_{\text{scatt}}=\pm18.70$ fm are not very far from the second order curve
for $b_{2}(T)=0.530$. \ However the results for $-18.70$ fm and $+18.70$ fm
both lie above the second order virial curve, rather than one on either side.

%

\begin{figure}
[ptb]
\begin{center}
\includegraphics[
height=3.6288in,
width=3.3935in,
angle=-90
]%
{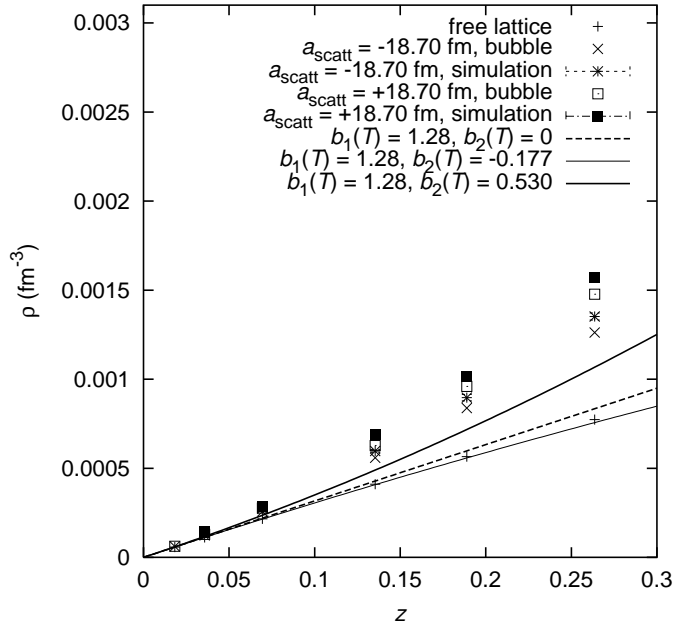}%
\caption{Density versus fugacity at $T=$ $3$\ MeV for scattering lengths
$a_{\text{scatt}}=\pm18.70$ fm. \ We show comparisons with the first and
second order virial expansion.}%
\label{rho_z_3}%
\end{center}
\end{figure}

\begin{figure}
[ptb]
\begin{center}
\includegraphics[
height=3.6288in,
width=3.3935in,
angle=-90
]%
{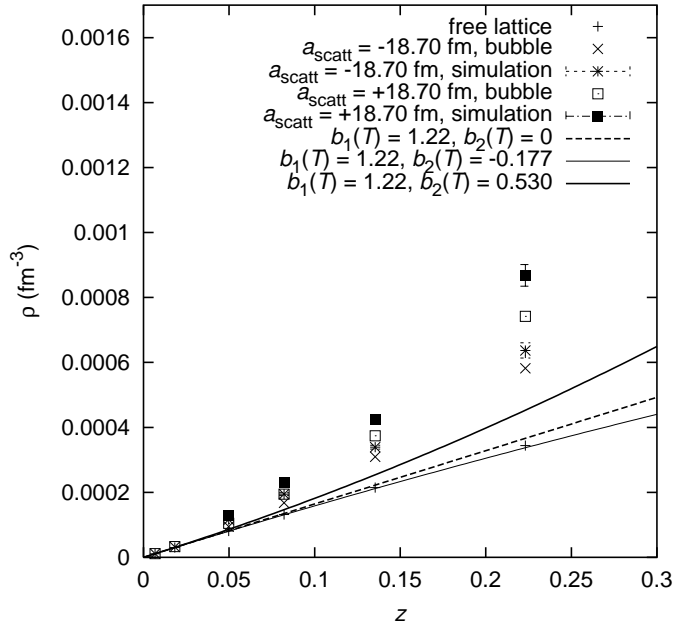}%
\caption{Density versus fugacity at $T=$ $2$\ MeV for scattering lengths
$a_{\text{scatt}}=\pm18.70$ fm. \ We show comparisons with the first and
second order virial expansion.}%
\label{rho_z_2}%
\end{center}
\end{figure}

In Fig. \ref{rho_z_3} we plot the density versus fugacity at $T=3$ MeV and
virial curves with $b_{1}(T)=1.28$. \ In Fig. \ref{rho_z_2} we plot the
density versus fugacity at $T=2$ MeV and virial curves with $b_{1}(T)=1.22$.
\ Again we determined $b_{1}(T)$ by fitting the bubble chain data at very
small fugacity. \ The free Fermi gas results on the lattice at $T=3$ MeV and
$2$ MeV match the second order virial results at $b_{2}(T)=-0.177$. \ But we
find the same problem for the bubble chain and full simulation results. \ The
results for $a_{\text{scatt}}=-18.70$ fm and $+18.70$ fm both lie above the
second order virial curve for $b_{2}(T)=0.530$. \ Furthermore the deviation
appears to be worse at colder temperatures.

%

%


\section{Lattice virial coefficients}

\label{sec_lvir}

We use the bubble chain expansion to compute $b_{2}(T)$ on the lattice for the
various coupling strengths and temperatures mentioned above. \ The specific
procedure we use is as follows. \ We first compute the free Fermi gas density
on the lattice and use the relation%
\begin{equation}
\rho^{\text{free}}\approx\frac{2}{\lambda_{T}^{3}}b_{1}(T)\left[
z-2\cdot2^{-\frac{5}{2}}z^{2}\right]
\end{equation}
to determine $b_{1}(T)$. \ This can be measured at any small fugacity, and we
choose $z=e^{-5}\approx0.0067$ or $\mu=-5k_{B}T$. \ In Table 2 we show results
for $b_{1}(T)$ for a range of temperatures. \ We have taken the lattice volume
to be large enough so that the finite volume error is $0.1\%$ or less, and the
values for $L$ are shown in Table 2.%
\[%
\genfrac{}{}{0pt}{0}{\text{Table 2: \ }b_{1}(T)\text{ on the lattice}}{%
\begin{tabular}
[c]{|l|l|l|}\hline
$T\text{ (MeV})$ & $L$ & $b_{1}(T)$\\\hline
$4$ & $5$ & $1.273$\\\hline
$3$ & $6$ & $1.282$\\\hline
$2$ & $6$ & $1.221$\\\hline
$1.5$ & $7$ & $1.160$\\\hline
$1$ & $8$ & $1.091$\\\hline
$0.667$ & $9$ & $1.053$\\\hline
$0.5$ & $10$ & $1.038$\\\hline
\end{tabular}
}%
\]
As the temperature decreases we see that $b_{1}(T)\rightarrow1$. \ This is
expected since at fixed lattice spacing the system moves closer to the
continuum limit as we decrease the temperature.

With the same chemical potential and lattice volumes, we compute the density
in the bubble chain approximation and determine $b_{2}(T)$ using the relation%
\begin{equation}
\rho^{\text{bubble}}\approx\frac{2}{\lambda_{T}^{3}}b_{1}(T)\left[  z+2\cdot
b_{2}(T)z^{2}\right]  \text{.}%
\end{equation}
The results for $b_{2}(T)$ for the various coupling strengths are shown in
Table 3. \ For each pair shown the lattice result is on the left and the
continuum result is on the right.%
\[%
\genfrac{}{}{0pt}{0}{%
\begin{array}
[c]{c}%
\text{Table 3: Second virial coefficient }b_{2}(T)\text{ for different
scattering lengths }a_{\text{scatt}}\text{ as }\\
\text{calculated on the lattice (first column) and in the continuum limit
(second column).}%
\end{array}
}{%
\begin{tabular}
[c]{|l||l|l||l|l||l|l||l|l||l|l||}\hline
$a_{\text{scatt}}$ (fm) & \multicolumn{2}{l|}{$-4.675$} &
\multicolumn{2}{l|}{$-9.35$} & \multicolumn{2}{l|}{$-18.70$} &
\multicolumn{2}{l|}{$+18.70$} & \multicolumn{2}{l|}{$+9.35$}\\\hline
$C$ (10$^{-4}$ MeV$^{-2}$) & \multicolumn{2}{l|}{$-1.028$} &
\multicolumn{2}{l|}{$-1.108$} & \multicolumn{2}{l|}{$-1.153$} &
\multicolumn{2}{l|}{$-1.257$} & \multicolumn{2}{l|}{$-1.318$}\\\hline
$T$ = 4 MeV & 0.66 & 0.198 & 0.88 & 0.322 & 1.03 & 0.411 & 1.45 & 0.692 &
1.77 & 0.917\\\hline
$T$ = 3 MeV & 0.76 & 0.170 & 1.07 & 0.299 & 1.29 & 0.396 & 1.97 & 0.722 &
2.52 & 1.004\\\hline
$T$ = 2 MeV & 0.80 & 0.131 & 1.26 & 0.263 & 1.61 & 0.371 & 2.85 & 0.776 &
3.98 & 1.176\\\hline
$T$ = 1.5 MeV & 0.71 & 0.103 & 1.23 & 0.237 & 1.68 & 0.352 & 3.42 & 0.825 &
5.22 & 1.350\\\hline
$T$ = 1 MeV & 0.46 & 0.066 & 1.09 & 0.198 & 1.44 & 0.322 & 3.79 & 0.917 &
6.85 & 1.720\\\hline
$T$ = 0.667 MeV & 0.23 & 0.031 & 0.60 & 0.159 & 1.01 & 0.289 & 3.58 & 1.047 &
8.01 & 2.368\\\hline
$T$ = 0.5 MeV & 0.12 & 0.008 & 0.40 & 0.131 & 0.74 & 0.263 & 3.29 & 1.176 &
8.82 & 3.167\\\hline
\end{tabular}
}%
\]
We see that the lattice virial coefficients are larger than the continuum
values. \ Also as we decrease the temperature at fixed scattering length, the
deviation in $b_{2}(T)$ first increases before eventually decreasing.


\section{Discussion}

\label{sec_dis}

\begin{figure}
[ptb]
\begin{center}
\includegraphics[
height=4.2436in,
width=2.981in,
angle=-90
]%
{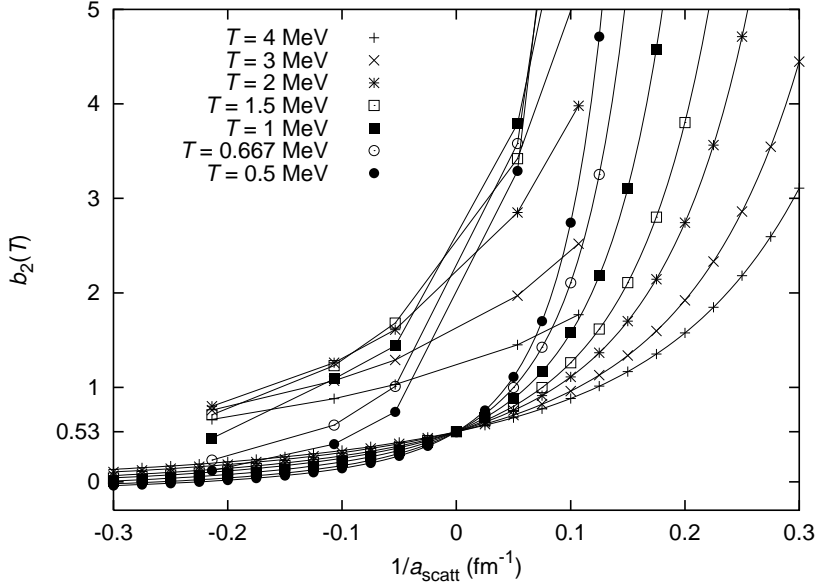}%
\caption{Plot of second virial coefficient as function of inverse scattering
length for various temperatures. \ The smooth curves are the continuum
results, and the points with interpolating lines are the lattice results.}%
\label{virialcompare}%
\end{center}
\end{figure}

\begin{figure}
[ptbptb]
\begin{center}
\includegraphics[
height=4.2436in,
width=2.981in,
angle=-90
]%
{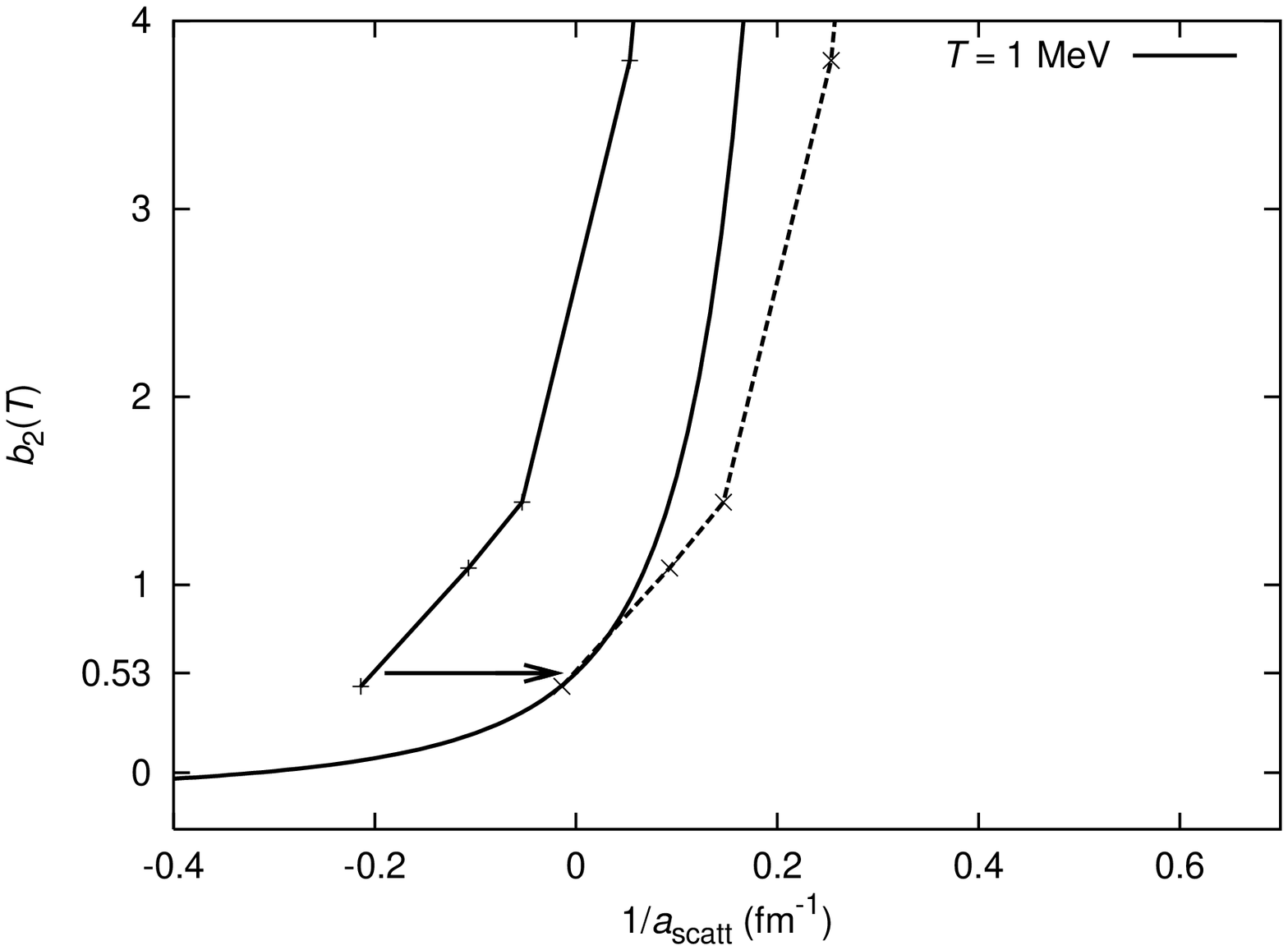}%
\caption{Plot of continuum and lattice virial results for $T=1$ MeV. \ We also
show the lattice results shifted horizontally so that $b_{2}$ equals $0.530$
at infinite scattering length.}%
\label{shifted_1}%
\end{center}
\end{figure}

\begin{figure}
[ptbptbptb]
\begin{center}
\includegraphics[
height=4.2436in,
width=2.981in,
angle=-90
]%
{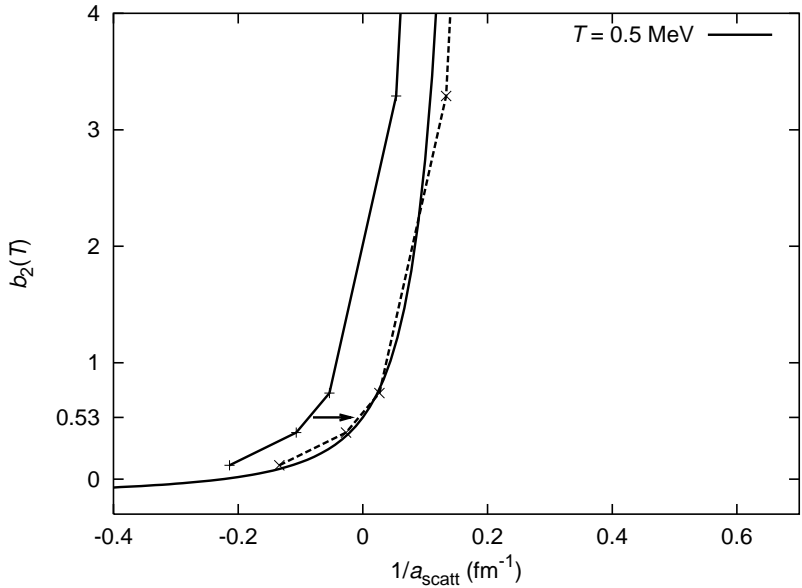}%
\caption{Plot of continuum and lattice virial results for $T=0.5$ MeV. \ We
also show the lattice results shifted horizontally so that $b_{2}$ equals
$0.530$ at infinite scattering length.}%
\label{shifted_0.5}%
\end{center}
\end{figure}

In Fig. \ref{virialcompare} we plot $b_{2}(T)$ versus inverse scattering
length for a range of temperatures. \ The smooth curves are the continuum
results given in (\ref{b2}), and the points with interpolating guide lines are
the lattice results from Table 3. \ For the continuum curves we see that the
value of $b_{2}$ at infinite scattering length remains fixed at $3\cdot
2^{-\frac{5}{2}}\approx0.530$. \ However as the temperature decreases, the
slope as a function of inverse scattering length steepens. \ The functional
dependence of $b_{2}$ on $a_{\text{scatt}}$ and $T$ is roughly%
\begin{equation}
b_{2}\sim e^{x^{2}}=e^{\frac{1}{a_{\text{scatt}}^{2}m_{N}T}}.
\end{equation}
%
We observe that the lattice virial curves are actually not very different from
the continuum virial curves. \ However they are shifted to the left slightly
as a function of the inverse scattering length. \ The large slope together
with this leftward shift is responsible for the large lattice virial
coefficients. \ This is seen even more clearly in Figs. \ref{shifted_1} and
\ref{shifted_0.5} where we show the continuum and lattice virial results at
temperatures $T=1$ MeV and $0.5$ MeV along with shifted lattice results so
that $b_{2}$ equals $0.530$ at infinite scattering length.
%
We see that the shifted curves appears to match the continuum results
relatively well. \ The required shift as a function of temperature appears to
decrease with temperature, though the exact dependence requires further study.

Physically the large error in the second virial coefficient is directly
related to singularities in the sum over particle-particle bubble diagrams due
to either true bound states for positive scattering length or zero-energy
resonances in the unitary limit.\ Although the singularities are completely
integrable at nonzero temperature, on the lattice the momentum integrals are
finite sums which can be very sensitive to small perturbations. \ The obvious
way to address this problem, aside from working at a much smaller lattice
spacing, is to use an improved lattice action, one that includes more than
simple nearest-neighbor hopping terms for the kinetic energy as well as higher
order corrections to the interaction. \ We are currently investigating
improved actions, but this is not an easy task and not all improved operators
maintain the positivity of the euclidean action. \ 

We propose that even with an unimproved action one can improve the scaling
behavior of the data by tuning the coefficient $C$ to give the correct
physical value of the second virial coeffcient $b_{2}(T)$ for each desired
simulation temperature. \ Since $b_{2}(T)$ on the lattice is easily calculated
this is not much of an added burden. \ This procedure coincides with the
standard approach of fixing the zero temperature scattering length in the
limit that the lattice spacing goes to zero or as $T\rightarrow0$ at fixed
lattice spacing.\ Fixing $b_{2}(T)$ will obviously improve the scaling
behavior at low density. \ Whether it will also improve the data in the
degenerate regime is not a priori clear and this question will be the focus of
our companion paper.

Acknowledgments: This work is supported in part by the US Department of Energy
grants DE-FG-88ER40388 (T.S.) and DE-FG02-04ER41335 (D.L.).

\bibliographystyle{apsrev}
\bibliography{NuclearMatter}

\end{document}